\documentclass[a4paper]{article}
\usepackage[noadjust]{cite}
\usepackage{graphicx}
\usepackage[left=2cm, right=2cm, top=2cm]{geometry} 

\graphicspath{{figs/}}
\usepackage{amsmath,amssymb}
\usepackage{relsize}
\usepackage{algorithm} 
\usepackage{algorithmic}

\newcommand\NoDo{\renewcommand\algorithmicdo{}}

\newcommand\NoThen{\renewcommand\algorithmicthen{}}

\DeclareMathOperator*{\argmin}{arg\,min}
\newcommand{\T}{^{^{_T}}}
\newcommand{\CHI}{\mathlarger{\mathlarger{\chi}}}

 \newtheorem{thm}{Theorem}[section]
 
 \newtheorem{lem}[thm]{Lemma}
 
 \newtheorem{defn}[thm]{Definition}

\begin{document}
%
\title{Switching divergences for spectral learning in blind speech dereverberation}
%
%
%

\author{
Francisco J. Ibarrola
\thanks{\footnotesize Instituto de Investigaci\'on en Se\~nales, Sistemas e Inteligencia Computacional, sinc(i), UNL, CONICET, FICH, Ciudad Universitaria, CC 217, Ruta Nac. 168, km 472.4, (3000) Santa Fe, Argentina. ({\tt fibarrola@sinc.unl.edu.ar}).}
\and Leandro E. Di Persia
$^\ast$
\and Ruben D. Spies
\thanks{\footnotesize Instituto de Matem\'{a}tica Aplicada del Litoral, IMAL,  UNL, CONICET, Centro Cient\'ifico Tecnol\'ogico CONICET Santa Fe, Colectora Ruta Nac. 168, km 472, Paraje ``El Pozo'', (3000), Santa Fe, Argentina and Departamento de Matem\'{a}tica, Facultad de Ingenier\'{\i}a Qu\'{\i}mica, Universidad Nacional del Litoral, Santa Fe, Argentina.}
}
\maketitle

\begin{abstract}
When recorded in an enclosed room, a sound signal will most certainly get affected by reverberation. This not only undermines audio quality, but also poses a problem for many human-machine interaction technologies that use speech as their input. In this work, a new blind, two-stage dereverberation approach based in a generalized $\beta$-divergence as a fidelity term over a non-negative representation is proposed. The first stage consists of learning the spectral structure of the signal solely from the observed spectrogram, while the second stage is devoted to model reverberation. Both steps are taken by minimizing a cost function in which the aim is put either in constructing a dictionary or a good representation by changing the divergence involved. In addition, an approach for finding an optimal fidelity parameter for dictionary learning is proposed. An algorithm for implementing the proposed method is described and tested against state-of-the-art methods. Results show improvements for both artificial reverberation and real recordings.
\end{abstract}


\section{Introduction}

%

Over the last years, with the technological advances and massive adoption of portable electronic devices with high computational capacity, the need for better human-machine interaction capabilities has emerged as a topic of interest. Since speech constitutes one of the most natural ways of human communication, trying to achieve a fluid interaction with machines by this mean has been the subject of much recent research. This need for improvement is inherent to a number of hot topics in the field of signal processing, including automatic translation systems (\cite{yun_multilingual_2014}), emotion and affective state recognition (\cite{vignolo_feature_2016}), digital personal assistants (\cite{sarikaya2016overview}), to name just a few, that require the use of speech as inputs.

One of the main difficulties within this context comes from the fact that when recorded in enclosed rooms, audio signals are affected by reverberant components due to reflections of the sound waves in the walls, floor and ceiling. This can severely degrade the quality of the recorded signals (particularly when the microphones are far away from the sources, \cite{tashev2009sound}), which in turn makes them unsuitable for direct use in certain speech applications (\cite{huang2001spoken}). The goal of this work is to produce a  dereverberation technique for removing or highly attenuating the reverberant components of a recorded signal in order to enhance its quality.

A speech dereverberation problem can be classified as ``blind'' whenever the available data consist only of the reverberant signal itself, or as ``supervised'' when information of the environment or the speakers is available. The problem can also be classified as single or multi-channel, depending on the number of microphones used for recording. In this work, we shall address the problem within a blind, single-channel setting, which is the most common in real-life problems, but also the most difficult, because of the scarce information.

Due to the characteristics of speech signals, most state-of-the-art methods deal with the dereverberation problem in a transformed domain, such as the one obtained by the Fan-Chirp Transform (see \cite{wisdom2014}) or the Short-Time Fourier Transform (STFT) (\cite{ibarrola2018bayesian}). Some of these methods make use of non-negative matrix factorization (NMF) or its variants, such as convolutive NMF (\cite{smaragdis2004}), along with Bayesian or penalization approaches. Although such methods have shown to produce satisfactory results, they often neglect the relation between frequency components, for which some authors (e.g. \cite{mohammadiha2015joint}) have proposed an NMF model in which a \textit{dictionary} is used for spectral modeling. The main problem with this kind of models within a blind setting has to do with the scarce available data. That is, the dictionary should be good for representing a clean signal, while learnt from a reverberant one.

This article begins by presenting a convolutive NMF reverberation representation that uses a dictionary for spectral modeling, and proposing a general form for a cost function with mixed penalization for characterizing the model. Different variants of that cost function are used for stating a two-stage method, where the first stage takes care of building a dictionary, while the second one is devoted to use such dictionary for getting an appropriate representation of the reverberation model. The main novelty of this work is that the process of learning the spectral structure (\emph{i.e.} the first stage) is not aimed to obtain an optimal representation of the reverberant signal.

\section{Reverberation Model}

Let $s, x, h:\mathbb{R}\rightarrow\mathbb{R}$, supported in $[0, \infty)$, denote the functions associated to the clean and reverberant signals, and the room impulse response (RIR), respectively. As it is customary, we make the assumption that reverberation is well represented by a Linear Time-Invariant (LTI) system, which can be written as
\begin{equation} \label{eq:cont-model}
x(t) = (h\ast s)(t),
\end{equation}
where ``$\ast$'' denotes convolution. The use of this representation is underlaid by the hypotheses that the source and microphone positions are fixed, and the non-linear components are small enough to be neglected.

As we previously mentioned, when dealing with speech signals, it often results convenient to work with time-frequency representations rather than in the time domain. Thus, we shall make use of the Short Time Fourier Transform (STFT).

\subsection{STFT-based reverberation model}

The STFT of a function $x$ can be defined as
\begin{equation}\nonumber
\mathbf{x}_k(t) \doteq \int_{-\infty}^{\infty}x(u)w(u-t)e^{-2\pi i u k}du,\;\;t,k\in\mathbb{R},
\end{equation}
where $w:\mathbb{R}\rightarrow\mathbb{R}^+_0$ is a prescribed even and compactly supported function such that $\|w\|_1 = 1$, called \emph{window}.

Naturally, in practice we work with discretized versions of the signals, denoted as $x[\cdot],$ $h[\cdot],$ $s[\cdot],$ and $w[\cdot]$. The corresponding discrete STFT can be defined as
\begin{equation}\nonumber
\mathbf{x}_k[n] \doteq \sum_{m =-\infty}^{\infty}x[m]w[m-n]e^{-2\pi i m k},
\end{equation}
where $n=1,\ldots, N,$ is a discrete time variable associated to the window locations, and $k=1,\ldots, K,$ denotes the frequency sub-band.
Similarly, we denote by $\mathbf{s}_k[n]$ and $\mathbf{h}_k[n]$ the STFTs of $s$ and $h$, respectively. A discrete approximation of  (\ref{eq:cont-model}) in the STFT domain is given by
\begin{equation}\label{eq:discconv}
\mathbf{x}_k[n] \approx \tilde{\mathbf{x}}_k[n] \doteq \sum_{m = 0}^{M-1} \mathbf{s}_k[n-m] \mathbf{h}_k[m],\;\;n,k\in\mathbb{N}.
\end{equation}
where $M$ is a given model parameter determined by the reverberation time. The model is built as in \cite{avargel2007system}, where the approximation in (\ref{eq:discconv}) holds due to the use of band-to-band only filters. The window locations are chosen so that the support of the observed signal is contained in the union of the supports of the windows, and $K$ as to reach up to half the sampling frequency.

Since phase angles on the STFT components have been shown to be highly sensitive to mild variations on the associated signal (\cite{yegnanarayana1998}), and within our blind setting we have no information about reverberation conditions, we proceed as in \cite{kameoka2009}, by treating the phase angles $\phi_k[m]$ of $\mathbf{h}_k[m]$ as random variables. Let us assume them to be \emph{i.i.d.} with uniform distribution in $[-\pi,\pi)$. Under this hypothesis, it can be shown (\cite{ibarrola2018bayesian}) that  the expected value of $|\tilde{\mathbf{x}}_k[t]|^2$ is given by
\begin{align}
E|\tilde{\mathbf{x}}_k[n]|^2 = \sum_{m} |\mathbf{s}_k[n-m]|^2 \,|\mathbf{h}_k[m]|^2. \nonumber
\end{align}

Note that the choice of $[-\pi,\pi)$ is arbitrary, since the equality holds for any $2\pi-$length interval. Finally, by defining $S_{k,n} \doteq |\mathbf{s}_k[n]|^2$, $H_{k,n} \doteq |\mathbf{h}_k[n]|^2$ and $X_{k,n} \doteq E|\tilde{\mathbf{x}}_k[n]|^2$, the convolutive NMF model reads
\begin{equation} \label{eq:mod-rep}
X_{k,n} = \sum_{m=0}^{M'} S_{k,n-m} H_{k,m},
\end{equation}
for $k = 1, \ldots, K,\; n = 1, \ldots,N.$ Here, $M'\doteq \min\{M-1,n-1\}$, so we can treat $X$, $S$ and $H$ as nonnegative matrices with elements $X_{k,n}$, $S_{k,n}$ and $H_{k,n}$, respectively.

Since we intend to introduce a spectral modeling of the clean signal, we shall make use of an NMF approach over the clean spectrogram $S$.

\subsection{NMF model}

Let us assume that there exist $W\in \mathbb{R}_{0,+}^{K\times J},\;\; U\in \mathbb{R}_{0,+}^{J\times N}$, ($J<\min\{K,N\}$) that provide a ``good'' NMF representation for $S\in \mathbb{R}_{0,+}^{K\times N}$. That is,
\begin{equation} \nonumber
S \cong WU.
\end{equation}
The accuracy of this approximation can be defined in terms of the Euclidean distance or some divergence measure (details on this will be discussed later on). In order to keep the notation simple, we shall assume the latter approximation to hold exactly and replace $S$ in (\ref{eq:mod-rep}) by $WU$, which results in the model
\begin{equation} \label{eq:mod-WUH}
X_{k,n} = \sum_{m = 0}^{M'}  \sum_{j = 1}^{J}  W_{k,j}U_{j,n-m} H_{k,m}.
\end{equation}

Two remarks are in order: firstly, note that the approximation error in the assumption $S=WU$ will be taken into account by the representation error of $X$ with respect to the data, and hence the latter assumption poses no problem. Secondly, we note that the model (\ref{eq:mod-WUH}) has a scale indeterminacy, in the sense that for any $\alpha >0$, the matrices $\tilde{W}=\alpha W$, $\tilde{H}=\alpha H$, and $\tilde{U}=\alpha^{-2}U$  would give the same representation $X$. Hence, in order to avoid numerical issues, we add the constraints $\|W_j\|_1 = \|H_k\T\|_\infty = 1$, where $W_j,\;j=1,\ldots, J,$ are the columns of $W$ and $H_k,\;k=1, \ldots, K$ are the rows of $H$. This means that the spectrogram $S$ is represented by a normalized dictionary and that reverberation preserves the  signal's maximal energy.

In the next section, a fidelity term and penalizers for building an appropriate cost function $f$ will be defined. This cost function will then be minimized in order to obtain the desired matrices $\hat{W}$, $\hat{U}$ and $\hat{H}$, as follows:

\emph{Algorithm overview}
\begin{enumerate}
\item Set the parameters of $f = f(Y,X)$ so as to prioritize spectral learning and minimize $f$ with respect to its arguments in order to find an appropriate dictionary $\hat{W}$.
\item  Reset the parameters of $f$ in order to emphasize accuracy in the representation. Then minimize $f$ with respect to $U$ and $H$ subject to $W=\hat{W}$, to obtain $\hat{U}$ and $\hat{H}$.
\item Approximate the clean spectrogram $S$ using $\hat{W}$ and $\hat{U}$.
\end{enumerate}

\section{Cost function}

\subsection{Fidelity term}

Given a reverberant (and possibly noisy) spectrogram $Y$, we intend to find matrices $W$, $U$ and $H$ that, while complying with certain desired characteristics, provide a representation $X$, as in (\ref{eq:mod-WUH}), that accurately approximates $Y$.

\begin{figure*} [!t]
	\centering
\includegraphics[width=1\textwidth]{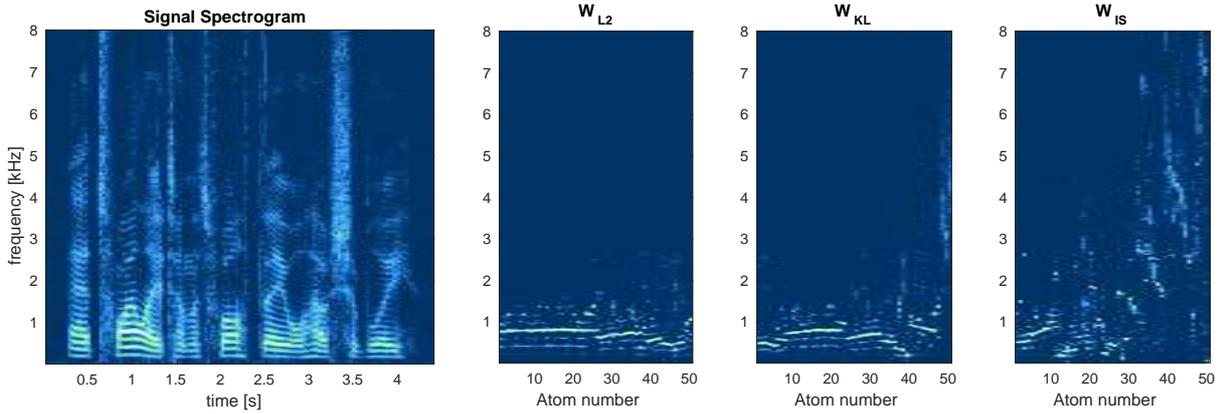}
\caption{Left: The spectrogram of a clean signal, sampled at 16[kHz], using a 512 samples window with overlapping of 256. $W_\text{L2}$: dictionary obtained using Frobenius norm. $W_\text{KL}$: dictionary obtained using Kullback-Leibler divergence. $W_\text{IS}$: dictionary obtained using Itakura-Saito divergence. All the dictionary atoms were ordered by correlation in order to help visualization.}
\label{fi:dictionaries}
\end{figure*}

Many ways of measuring the fidelity of that approximation have been proposed: the Euclidean distance (\cite{kameoka2009}), the Kullback-Leibler divergence (\cite{mohammadiha2015joint}), and the Itakura-Saito divergence (\cite{fevotte2009nonnegative}) being the most commonly used. Assume we have a known clean spectrogram $S$ that we want to represent using an NMF factorization $WU$. Different choices of the fidelity measure will lead to dictionary atoms (column vectors of $W$) with different characteristics. As it can be seen in Fig. \ref{fi:dictionaries}, a particular fidelity measure may emphasize the appearance of atoms that enable a good approximation in the higher energy zones while neglecting the low-energy ones, while another fidelity measure may result in the opposite. 

In order to find an ``optimal'' dictionary $W$, we begin by recalling a generalized divergence, as introduced in \cite{kompass2007generalized}. For $X, Y \in \mathbb{R}^{K\times N}_{0,+}$ and $\beta \in\mathbb{R}_+\backslash\{1\}$, the $\beta$-divergence of $X$ from $Y$ is defined as
\begin{align}
D_{\beta}(Y||X) \doteq & \sum_{k,n}\left( Y_{k,n}\frac{Y_{k,n}^{\beta-1}-X_{k,n}^{\beta-1}}{\beta(\beta-1)} +X_{k,n}^{\beta-1}\frac{X_{k,n}-Y_{k,n}}{\beta}\right). \nonumber
\end{align}

This $\beta$-divergence generalizes all three aforementioned fidelity measures. In fact, it can be seen that $D_{2}(\cdot||\cdot)$  corresponds to (half) the squared Frobenius norm of $Y-X$, whereas $D_{\beta}(\cdot||\cdot)$  approaches the Kullback-Leibler divergence as $\beta\rightarrow1$ and the Itakura-Saito divergence as $\beta\rightarrow 0$. An appropriate way of choosing the parameter $\beta$ will be discussed later on. We now proceed to introduce the penalization terms which shall embed the desired characteristics on the components that constitute the model.

\subsection{Penalizers}

Clearly, there are many ways of building the matrices $W, U$ and $H$ leading to a representation with small divergence with respect to the observation. One way of narrowing down the possible choices is by introducing penalizing terms into our cost function for promoting certain desired features over its minimizers. In a quite general context, this leads to a cost function of the form
\begin{equation}
f(W,U,H) \doteq D_{\beta}(Y||X) +P_u(U) +P_h(H), \nonumber
\end{equation}
where $P_u:\mathbb{R}_{0,+}^{J\times N} \rightarrow \mathbb{R}_{0,+}$, and $P_h:\mathbb{R}_{0,+}^{K\times M} \rightarrow \mathbb{R}_{0,+}$ are penalizing functions, each one imposing a cost over the appearance of certain features on $U$ and $H$, respectively.

\begin{figure} [h]
	\centering
\includegraphics[width=0.45\textwidth]{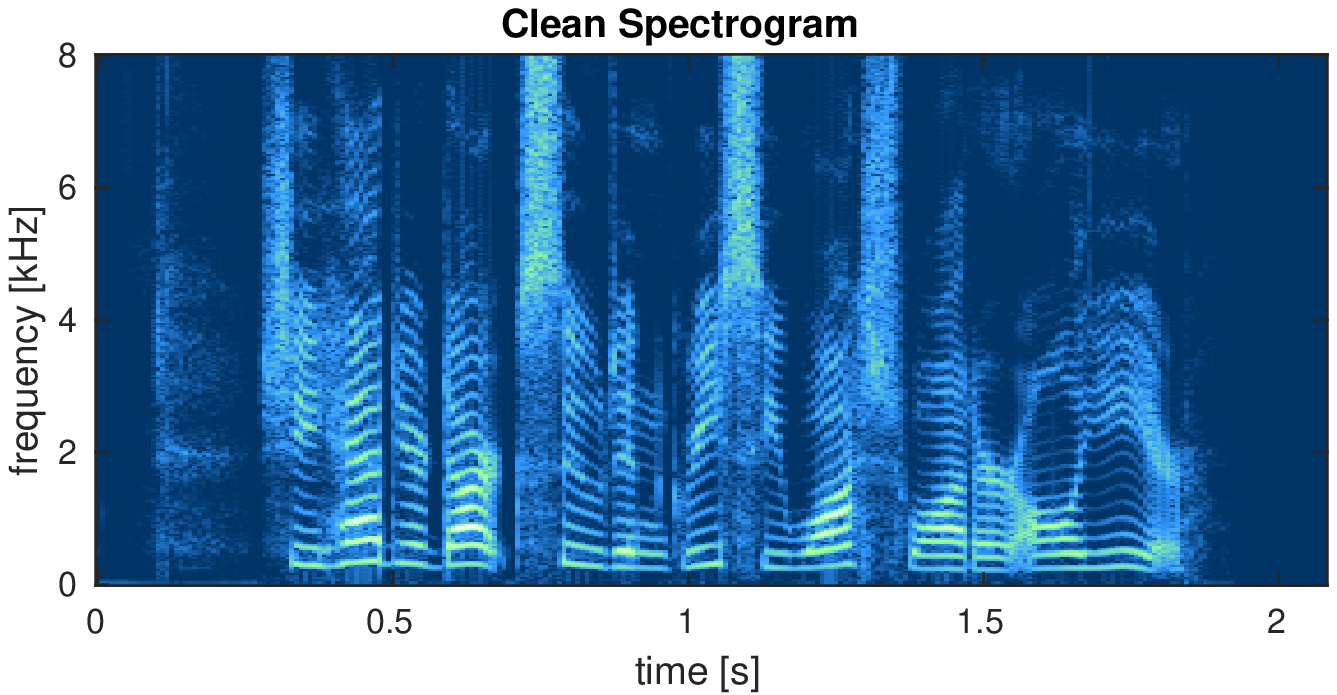}

\includegraphics[width=0.45\textwidth]{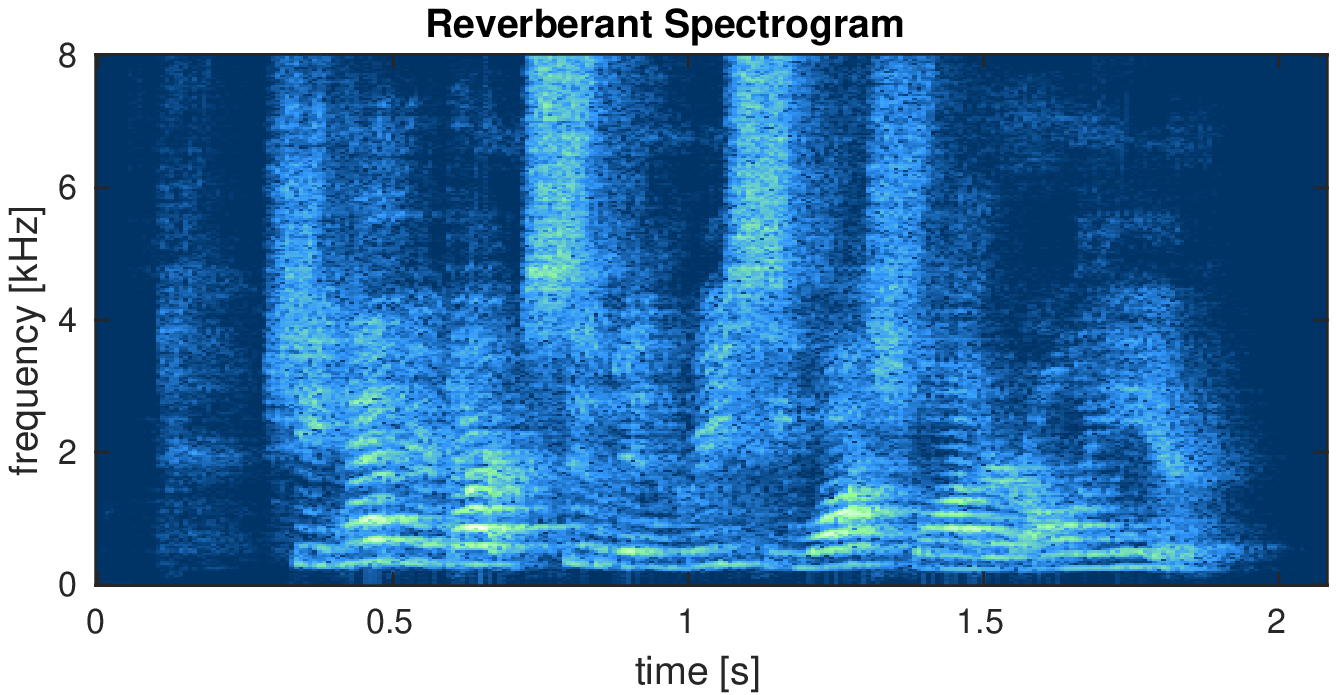}
\caption{Top: spectrogram of a clean signal, sampled at 16[kHz], using a 512 samples window with overlapping of 256. Bottom: the spectrogram of a reverberant ($600$[ms]) version of the same signal.}
\label{fi:clean_vs_rev}
\end{figure}

As it can be observed, while the spectrogram of the clean signal depicted in Fig. \ref{fi:clean_vs_rev} presents a somewhat sparse structure, the one corresponding to the reverberant signal  presents a smoother, more diffuse structure. As it is customary (\cite{mohammadiha2015joint}), we shall hinder the smoothness observed in the reverberant spectrogram from appearing in the restored spectrogram by defining a penalizer over the activation coefficients matrix $U$ of the form
\begin{equation}
P_u(U) \doteq \sum_{j,n}  \lambda_{n}^{(u)} U_{j,n},\nonumber
\end{equation}
where $ \lambda_{n}^{(u)}\geq 0$, $ n = 1, \ldots, N,$  are called penalization parameters for $P_u$. We let the penalizer depend on the time index $n$ as to allow for better compliance with the inherent silences of the recorded signals (more on this subject in Section \ref{sec:recording_experiments}).

\begin{figure} [!t]
	\centering
\includegraphics[width=0.45\textwidth]{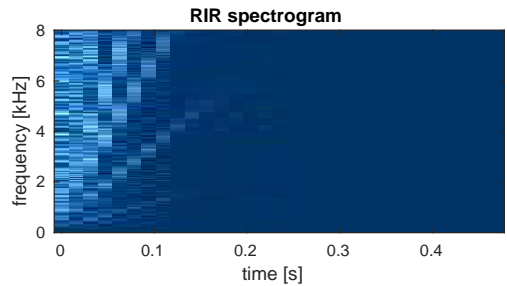}
\caption{Log-spectrogram for an artificial 16 [kHz] RIR signal with reverberation time of 450 [ms]. The spectrogram was made using a Hanning window length of 512 and overlapping of 256.}
\label{fi:rirspecs}
\end{figure}

In order to define a penalizer over $H$, we turn our attention to Fig. \ref{fi:rirspecs}, that shows a simulated RIR in a room with a reverberation time of $450$[ms]. The log-spectrogram exhibits a high-energy vertical band on the left, corresponding to the first echoes to reach the receiver, that slowly fades to the right, as deemed by a linear impulse response. The oblique straight lines of less energy correspond to an apparent frequency increase due to the increasing rate at which echoes reach the microphone in rectangular rooms (\cite{de2015modeling}). From these characteristics, and the fact that the overlapping of windows results in consecutive time components of $H$ capturing common information, it is reasonable to expect the components of $H$ to exhibit a smooth decay over time (\cite{ratnam2003blind}). This structure can be promoted (see \cite{ibarrola2018bayesian}) by introducing a penalizer of the form
\begin{equation}
P_h(H) \doteq \sum_{k}\lambda^{(h)}_{k} \|LH_k\T\|_2^2,\nonumber
\end{equation}
where $\lambda_{k}^{(h)}\geq0$, $H_k\in\mathbb{R}_{0,+}^M,\; k=1,\ldots , K$ are the rows of $H$, and $L\in\mathbb{R}^{(M-1)\times M}$ is a \textit{finite difference matrix}, so that $[LH_k\T]_m = H_{k,m+1}-H_{k,m}$.

With all of the above, the cost function is defined as follows:
\begin{equation} \label{eq:functional}
f(W,U,H) \doteq D_{\beta}(Y||X) + \sum_{j,n}  \lambda_{n}^{(u)} U_{j,n}
+\sum_{k}\lambda_{k}^{(h)} \|LH_k\T\|_2^2.
\end{equation}

In the next section we state a two-stage optimization process in order to minimize $f$, first with respect to $W$, and then with respect to both $U$ and $H$. In-line with the core idea stated before, by appropriately tunning its parameters, the cost function (\ref{eq:functional}) can be used for building a good dictionary in a first stage, and for seeking a good representation of the data in a second step.

\section{Optimization}

The optimization process that shall yield the restored spectrogram $\hat{S}$ is divided in two main steps: firstly, given the observed reverberant spectrogram $Y\in \mathbb{R}_{0,+}^{K\times N},$ a suitable dictionary $\hat{W}\in\mathbb{R}_{0,+}^{K\times J}$ that be able to provide a good representation of the target clean spectrogram $S$ is built. Once this is accomplished, the algorithm proceeds to find $\hat{U}\in \mathbb{R}_{0,+}^{J\times N}$ and $\hat{H}\in \mathbb{R}_{0,+}^{K\times M}$ minimizing $f$ given $\hat{W}$.

In order to minimize the cost function, we shall begin by introducing the concept of auxiliary function.

\subsection{Auxiliary function}

\begin{defn}\label{def:aux_fun}
Let $\Omega\subset\mathbb{R}^P$ and $f:\Omega\rightarrow\mathbb{R}_0^+$. Then, $g:\Omega\times\Omega\rightarrow\mathbb{R}_0^+$ is called an \emph{auxiliary function} for $f$ if $ g(\omega,\omega) = f(\omega)$ and $g(\omega,\omega')\geq f(\omega), \;\; \forall \omega,\omega'\in\Omega.$
\end{defn}

\begin{lem}
\label{lem:aux_fun}
If we let $f$ and $g$ be as in the definition above, $\omega^0\in\Omega$ be arbitrary and
\begin{equation} \nonumber 
\omega^t \doteq \argmin_\omega g(\omega,\omega^{t-1}),\; t\in\mathbb{N}
\end{equation}
then it can be shown (\cite{lee2001}) that the sequence $\{f(\omega^t)\}_{t\geq1}$ is non-increasing.
\end{lem}

The idea is to build an auxiliary function $g$ for $f$ with respect to each of its three arguments individually, and then use them iteratively for minimizing $f$.

We will proceed in a similar fashion than in \cite{fevotte2011algorithms}. Firstly, let us notice that $\forall Y\in\mathbb{R}^{K \times N}_{0,+}, \;D_\beta(Y||\,\cdot\,) \in \mathcal{C}^\infty(\mathbb{R}_+^{K\times N})$, and
\begin{equation}\label{eq:ddD}
\frac{\partial^2 D_\beta(Y||X)}{\partial X_{k,n}^2} = (\beta-1)X_{k,n}^{\beta-2}+(2-\beta)X_{k,n}^{\beta-3}Y_{k,n}.
\end{equation}
By defining
\begin{align*}
\check{D}_\beta(Y||X)  & \doteq \sum_{k,n}\left( \frac{\CHI_{\beta>1}(\beta)}{\beta}X_{k,n}^{\beta} -\frac{\CHI_{\beta\leq 2}(\beta)}{\beta -1}Y_{k,n} X^{\beta-1}_{k,n}+\frac{1}{\beta(\beta -1)}Y^{\beta}_{k,n}\right), 
\end{align*}
and
\begin{align*}
\hat{D}_\beta(Y||X) & \doteq \sum_{k,n}\left(\frac{\CHI_{\beta<1}(\beta)}{\beta}X_{k,n}^{\beta} -\frac{\CHI_{\beta > 2}(\beta)}{\beta -1}Y_{k,n} X^{\beta-1}_{k,n}\right), 
\end{align*}
we have $D_\beta = \check{D}_\beta + \hat{D}_\beta$, where $\check{D}_\beta$ is convex and  $\hat{D}_\beta$ is concave (both w.r.t. $X$). In the following, we will make use of this decomposition in order to build auxiliary functions for updating each one of the components of $X$.

\subsection{Building $\hat{W}$}

As mentioned before, the parameters required for building a proper dictionary $\hat{W}$ are not necessarily the same as those leading to an optimal representation. Thus, we begin by fixing $H_{k,n} = 1$ if $n=1$ and $H_{k,n}=0,\forall n = 2, \ldots, M,\;\; k =1 \ldots, K.$ This means that we are precluding $H$ from modeling reverberation, and henceforth it does not make sense to promote temporal sparsity over $U$, and so we set $\lambda_{n}^{(u)}=0,\; \forall n  = 1, \ldots, N$, only for the first stage.

Now, provided we have found adequate parameters (what we address in Section \ref{sec:recording_experiments}), the problem of finding an appropriate dictionary reduces to minimizing (\ref{eq:functional}) with respect to $W$ and $U$ subject to $H$ and $\lambda_n^{(u)}$ be set as above. To do so, we begin by finding an auxiliary function for (\ref{eq:functional}) w.r.t. $W$. Let $W' \in\mathbb{R}_+^{K\times J}$, and let us denote $X_{k,n}' = \sum_{j,m} W'_{k,j} U_{j,n-m}H_{k,m}$. Then,
\begin{align}
\check{D}_\beta(Y_{k,n}||X_{k,n}) & = \check{D}_\beta\left( Y_{k,n} \bigg|\bigg| \sum_{j,m} W_{k,j} U_{j,n-m}H_{k,m} \right) \nonumber \\
& = \check{D}_\beta\left( Y_{k,n} \bigg|\bigg| \frac{\sum_{j,m} W_{k,j} U_{j,n-m}H_{k,m} X'_{k,n} \frac{W'_{k,j}}{W'_{k,j}}}{X'_{k,n}}\right) \nonumber\\
& = \check{D}_\beta\left( Y_{k,n} \bigg|\bigg| \frac{\sum_{j,m} W'_{k,j} U_{j,n-m}H_{k,m} X'_{k,n} \frac{W_{k,j}}{W'_{k,j}}}{\sum_{j,m} W'_{k,j} U_{j,n-m}H_{k,m}}\right) \nonumber\\
& \leq \sum_{j,m} \frac{W'_{k,j} U_{j,n-m}H_{k,m} }{X'_{k,n}}\check{D}_\beta\left( Y_{k,n} \bigg|\bigg| X'_{k,n}\frac{W_{k,j}}{W'_{k,j}}\right), \label{eq:ineq_convex} 
\end{align} 
where the last step is due to Jensen's inequality.

In regard to $\hat{D}_\beta$, since it is concave w.r.t. $X$, it follows that
\begin{align}
\hat{D}_\beta(Y_{k,n}||X_{k,n}) \leq \hat{D}_\beta(Y_{k,n}||X'_{k,n}) \label{eq:ineq_concave}+\frac{\partial \hat{D}_\beta(Y_{k,n}||X'_{k,n})}{\partial X_{k,n}}\sum_{j,m} (W_{k,j}-W'_{k,j}) U_{j,n-m}H_{k,m}.
\end{align}

Given $U$ and $H$ fixed, let us define $g_w:\mathbb{R}_+^{K\times J} \times \mathbb{R}_+^{K\times J} \rightarrow \mathbb{R}$ by
\begin{align*}
g_w(W,W')  & \doteq  \sum_{k,n,j,m}  \frac{W'_{k,j} U_{j,n-m}H_{k,m} }{X'_{k,n}}\check{D}_\beta\left( Y_{k,n}\bigg|\bigg| X'_{k,n}\frac{W_{k,j}}{W'_{k,j}}\right) \\
&+  \sum_{k,n,j,m}  \frac{\partial \hat{D}_\beta(Y_{k,n}||X'_{k,n})}{\partial X_{k,n}}(W_{k,j}-W'_{k,j}) U_{j,n-m}H_{k,m}\\
&+ \sum_{k,n} \hat{D}_\beta(Y_{k,n}||X'_{k,n}).
\end{align*}
Then, it follows from (\ref{eq:ineq_convex}) and (\ref{eq:ineq_concave}) that $g_w$ is an auxiliary function for $f$ w.r.t. $H$. Note that the equality condition in Definition \ref{def:aux_fun} also holds.

Since $g_w(W,W')$ is convex with respect to $W$, it can be minimized by equating its gradient to zero, what leads to
\begin{align*} 
0 = \left(\frac{W_{k,j}}{W'_{k,j}}\right)^{\alpha_1} \sum_{n,m}X_{k,n}'^{\beta-1}U_{j,m}H_{k,n-m} -\left(\frac{W_{k,j}}{W'_{k,j}}\right)^{\alpha_2} \sum_{n,m}X_{k,n}'^{\beta-2}Y_{k,n}U_{j,m}H_{k,n-m},
\end{align*}
where $\alpha_1 = (\beta-1) \CHI_{\beta>1}(\beta)$, and $\alpha_2 = (\beta-2) \CHI_{\beta\leq 2}(\beta)$. This automatically leads to the updating equation
\begin{equation}\label{eq:wupdate}
W^{(t)}_{k,j} = W^{(t-1)}_{k,j} \frac{\left[\left(\sum\limits_{m,n}\left(X^{(t-1)}_{k,n}\right)^{\beta-2}Y_{k,n}U_{j,m}H_{k,n-m} \right)^\eta \right]_{\epsilon}}{\left( \sum\limits_{m,n}\left(X^{(t-1)}_{k,n}\right)^{\beta-1}U_{j,m}H_{k,n-m} \right)^\eta},
\end{equation}
where $\eta \doteq \frac{1}{\alpha_1-\alpha_2}$. Here, the supra index $t$ denotes the iteration number and $[\cdot]_{\epsilon}$ denotes the operation $\max\{\cdot\,,\epsilon\}$ , with $\epsilon$ being a small constant ($\sim 10^{-10}$). This is used to avoid the elements of $W$ from dropping to 0 (or below), as once an element is null, it cannot regain positive values by a multiplicative updating procedure (see \cite{choi2005blind}). For simplicity of notation, we have avoided the use of superscripts in all the variables that do not depend directly on $W$.

In a similar fashion, it can be shown that an auxiliary function for $f$ with respect to $U$ is given by
\begin{align*}
g_u(U,U')  & \doteq    \sum_{k,n,j,m}    \frac{W_{k,j} U'_{j,m}H_{k,n-m} }{X'_{k,n}}\check{D}_\beta\left( Y_{k,n} \bigg|\bigg| X'_{k,n}\frac{U_{j,m}}{U'_{j,m}}\right)   \\
&+   \sum_{k,n,j,m}    \frac{\partial \hat{D}_\beta(Y_{k,n}||X'_{k,n})}{\partial X_{k,n}}W_{k,j}(U_{j,m}-U'_{j,m}) H_{k,n- m} \\
&+\sum_{k,n}\hat{D}_\beta(Y_{k,n}||X'_{k,n})+\sum_{j,n}\lambda^{(u)}_n U_{j,n}.
\end{align*}

Here again, since $g_u(U,\cdot)$ is convex, it can be minimized by equating its gradient to zero, which is tantamount to solving
\begin{equation} \nonumber
U_{j,m}  =  U'_{j,m}
\left( \frac{\sum\limits_{k,n}X'^{\beta-2}_{k,n}Y_{k,n}W_{k,j}H_{k,n-m} -\lambda^{(u)}_m \left( \frac{U'_{j,m}}{U_{j,m}}\right)^{\alpha_2} }
{\sum\limits_{k,n}X'^{\beta-1}_{k,n}W_{k,j}H_{k,n-m}}\right)^{\eta} .
\end{equation}

Let us notice that this is an implicit equation with respect to $U_{j,m}$ for $\beta<2$ (and $\lambda^{(u)}_j \neq 0$), but since $g_u$ is an auxiliary function for $f$ w.r.t. $U$, Lemma \ref{lem:aux_fun} guarantees that $U^{(t)}$ approaches a limit $\hat{U}$ as $t$ tends to infinity, and so the quotient $U_{j,m}^{(t)}/U_{j,m}^{(t-1)}$ should approach 1. Henceforth, the approximation $U_{j,m}^{(t)}/U_{j,m}^{(t-1)}\approx 1$ yields the following multiplicative updating rule:
\begin{equation} \label{eq:uupdate}
U^{(t)}_{j,m}  =  U^{(t-1)}_{j,m}
 \frac{\left[\left(\sum\limits_ {k,n}\left(X_{k,n}^{(t-1)}\right)^{ \beta-2} Y_{k,n}W_{k,j}H_{k,n-j} -\lambda^{(u)}_m\right)^{\eta} \right]_{\epsilon}}
{\left(\sum\limits_{k,n}\left(X_{k,n}^{(t-1)}\right)^{\beta-1}W_{k,j}H_{k,n-j}\right)^{\eta} } .
\end{equation}

The dictionary $\hat{W} = \argmin_{W}f(W,U,H)$ can thus be obtained by alternatively updating $W$ and $U$ using (\ref{eq:wupdate}) and (\ref{eq:uupdate}), respectively, until convergence.

Once $\hat{W}$ is obtained, we proceed to find $\hat{U}$ and $\hat{H}$ that be able to effectively model reverberation.

\subsection{Building $\hat{U}$ and $\hat{H}$}

Unlike in the first step, now we do want to impose a sparse structure over $U$, and so $\lambda^{(u)}_{n}$ should no longer be null for every $n = 1, \ldots, N$. Furthermore, it should be pointed out that the value of $\beta$ in this stage is not necessarily the same as in the previous one (and in fact they will be chosen differently in practice).

The updating rule for $U$ is exactly the same as stated in (\ref{eq:uupdate}). In regard to $H$, we define the auxiliary function %
\begin{align*}
g_h(H,H')& \doteq  \sum_{k,n,j,m}  \frac{W_{k,j} U_{j,n-m}H'_{k,m} }{X'_{k,n}}\check{D}_\beta\left( Y_{k,n}\bigg|\bigg| X'_{k,n}\frac{H_{k,m}}{H'_{k,m}}\right)   \\
&+\sum_{k,n,j,m}  \frac{\partial \hat{D}_{\beta}(Y_{k,n}||X'_{k,n})}{\partial X_{k,n}}(H_{k,m}-H'_{k,m})W_{k,j}U_{j,n-m}\\
&+\sum_{k,n}\hat{D}_\beta(Y_{k,n}||X'_{k,n}) +\sum_k\lambda^{(h)}_k\|LH_k\T\|^2.
\end{align*}
By equating its gradient (with respect to $H_{k,m}$) to zero, we obtain, for every $k = 1,\ldots, K, m = 1,\ldots, M,$
\begin{align*}
0 =& \sum_{j,n}W_{k,j} U_{j,n-m}\left(X'_{k,n}\right)^{\alpha_1}\left(\frac{H_{k,m}}{H'_{k,m}}\right)^{\alpha_1}
-\sum_{j,n}W_{k,j}U_{j,n-m}Y_{k,n}\left(X'_{k,n}\right)^{\alpha_2}\left(\frac{H_{k,m}}{H'_{k,m}}\right)^{\alpha_2}\\
&-2\lambda^{(h)}_k [L\T L H_k\T ]_m.
\end{align*}
It has been observed that using a multiplicative updating rule analogous to those used for $W^{(t)}$ and $U^{(t)}$ usually results in undesired oscillations in the elements of $H^{(t)}$. This is most likely due to the alternating signs in the rows of $L\T L$. In order to overcome this potential drawback, for every $k = 1, \ldots, K$, we define the diagonal matrix $A^{(k)}\in\mathbb{R}^{M\times M}_{0,+}$ with $A^{(k)}_{m,m} = \sum_{j,n}W_{k,j} U_{j,n-m}\left(X_{k,n}^{(t-1)}\right)^{\alpha_1}/H^{(t-1)}_{k,m}$ and define the vector $b^{(k)} \in\mathbb{R}^{M}_{0,+}$ as $b^{(k)} = \sum_{j,n}W_{k,j}U_{j,n-m}Y_{k,n}\left(X_{k,n}^{(t-1)}\right)^{\alpha_2}$. Then, under the same approximation used for arriving at (\ref{eq:uupdate}), we can update $H$ by solving for $H^{(t)}_{k},$ $k = 1, \ldots, K$, the linear system
\begin{equation} \label{eq:hupdate}
\left(A^{(k)} + 2\lambda_{k}^{(h)}L\T L\right)H^{(t)}_{k} = b^{(k)}.
\end{equation}
It can be shown that the matrix $A^{(k)} + 2\lambda_{k}^{(h)}L\T L$ is strictly positive definite (unless $A^{(k)}$ is null), and hence the linear system (\ref{eq:hupdate}) has a unique solution, whose elements are non-negative.

\subsection{Additional considerations}

Our approximate solution could be defined simply as $\hat{S} = \hat{W}\hat{U}$, but although this clearly leaves out reverberation (which is captured by $\hat{H}$), this low-rank approximation still entails some error. In order to avoid this, we estimate the clean spectrogram by multiplying the data elements $Y_{k,n}$ by a time-varying gain function $G_{k,n} \doteq  \frac{\sum_j \hat{W}_{k,j}\hat{U}_{j,n}}{\sum_{j,m} \hat{W}_{k,j}\hat{U}_{j,n-m},\hat{H}_{k,m}}$, as suggested in \cite{mohammadiha2015joint}.

All steps necessary for our dereverberation method are summarized in Algorithm \ref{al:beta_der}.\footnote{To try online: http://sinc.unl.edu.ar/web-demo/beta-dereverberation/}

\begin{algorithm}
\caption{Variable $\beta$-divergence dereverberation}
\label{al:beta_der}
\begin{algorithmic}
\NoDo
\NoThen
\item[\textbf{Preliminaries}]
\STATE Given a speech signal $y$, build $Y_{k,n} = |\text{STFT}(y)_{k,n}|^2$.

\vspace{0.2cm}
\item[\textbf{Stage 1}]

\STATE Set $\beta = \beta_1$ and $\lambda_{n}^{(u)}=0,\; \forall n.$
\STATE Let $H_{k,n} = 1$ if $n=1$ and $H_{k,n}=0,\forall n \geq 2, \forall k.$
\STATE Initialize $W^{(0)}$ and $U^{(0)}$ randomly.
\STATE Let $t=0$,
\WHILE{$\|W^{(t)} - W^{(t-1)}\|_F^2 >\delta$}
\STATE $t\leftarrow t+1$
\STATE Update $W^{(t)}$ as stated in (\ref{eq:wupdate}).
\STATE Update $U^{(t)}$  as stated in (\ref{eq:uupdate}).
\ENDWHILE
\STATE Let $\hat{W} = W^{(t)}$
\vspace{0.2cm}

\item[\textbf{Stage 2}]

\STATE Set $\beta = \beta_2$ and reset $\lambda_{n}^{(u)}\; \forall n.$
\STATE Let $H^{(0)}_{k,n} = \exp{(1-n)},$ $\forall n, k.$
\STATE Initialize $U^{(0)}$ as the last approximation in Stage 1.
\STATE Let $t=0$,
\WHILE{$\|S^{(t)}-S^{(t-1)}\|_F^2>\delta$}
\STATE $t\leftarrow t+1$
\STATE Update $U^{(t)}$ as stated in (\ref{eq:uupdate}).
\STATE Update $H^{(t)}$  as stated in (\ref{eq:hupdate}).
\ENDWHILE
\STATE Let $\hat{U} = U^{(t)}$
\STATE Let $\hat{H} = H^{(t)}$
\vspace{0.2cm}

\item[\textbf{Reconstruction}]

\STATE Let $G_{k,n} \doteq  \sum_j \hat{W}_{k,j}\hat{U}_{j,n}/\left(\sum_{j,m} \hat{W}_{k,j}\hat{U}_{j,n-m},\hat{H}_{k,m}\right)$.
\STATE Let $\hat{S}_{k,n} = G_{k,n}Y_{k,n}$.
\STATE Define $Z\in \mathbb{C}^{K\times N}$ by $Z_{k,n} = \sqrt{\hat{S}_{k,n}}\arg(Y_{k,n})$.
\STATE Define the restored signal in the time domain as \\ $\hat{s} \doteq \text{ISTFT}(Z)$.
\end{algorithmic}
\end{algorithm}

Next, we proceed to show some experimental results.

\section{Experimental results}

In this section we present a series of experiments, firstly for parameter search and then for validating our method. All signals used in the experiments were taken from the TIMIT database (\cite{zue1990timit}), sampled at $16$[kHz]. For the artificial RIR signals we made use of the software Room Impulse Response Generator\footnote{https://github.com/ehabets/RIR-Generator}.

In order to measure the quality of the restored signals, we used the well known frequency weighted segmental signal-to-noise ratio (fwsSNR) and the cepstral distance (\cite{hu2008}).  Additionally, we have computed the values of the speech-to-reverberation modulation energy ratio (SRMR, \cite{falk2010}). However, since the SRMR is non intrusive, its values must be used carefully for comparison purposes, keeping in mind that the resemblance of a restoration with the corresponding clean signal is not taken into account.

\subsection{Parameter estimation}

We begin by addressing the main parameter estimation problem for Stage 1 of Algorithm \ref{al:beta_der}. Namely, finding an optimal value of $\beta$ for building a dictionary whose atoms (columns) be able to provide a good representation of a clean spectrogram. In order to evaluate whether a given parameter $\beta_1$ is good for dictionary building, we take a reverberant spectrogram $Y$, build a dictionary $W^{(\beta_1)}$ by minimizing $D_{\beta_1}(Y||WU)$, and then proceed to check how well can $W^{(\beta_1)}$ represent the corresponding clean spectrogram $S$. To do this, given $\beta^*$, we minimize $D_{\beta^*}(S||W^{(\beta_1)}U)$ with respect to $U$. It is important to point out that in this second step, $\beta^*$ is not necessarily the same as $\beta_1$, and hence the two steps above are performed for every pair $(\beta_1,\beta^*)$ in order to find the optimal one.

To do this, we have taken five random clean signals and made them reverberant by means of a discrete convolution with an artificial RIR. For each reverberant spectrogram $Y$ and each admissible pair $(\beta_1,\beta^*)$, we have taken the following steps:
\begin{enumerate}
\item Build a dictionary $W^{(\beta_1)} = \argmin_{W,U} D_{\beta_1}(Y||WU)$.
\item Use $W^{(\beta_1)}$ to find a representation $\hat{S} = W^{(\beta_1)}\hat{U}$ for the associated clean spectrogram $S$, where $\hat{U} = \argmin_{U} D_{\beta^*}(S||W^{(\beta_1)}U)$.
\item Test the accuracy of the representation $\hat{S}$ by computing the cepstral distance with respect to $S$.
\end{enumerate}

\begin{figure} [!t]
\centering
\includegraphics[width=0.47\textwidth]{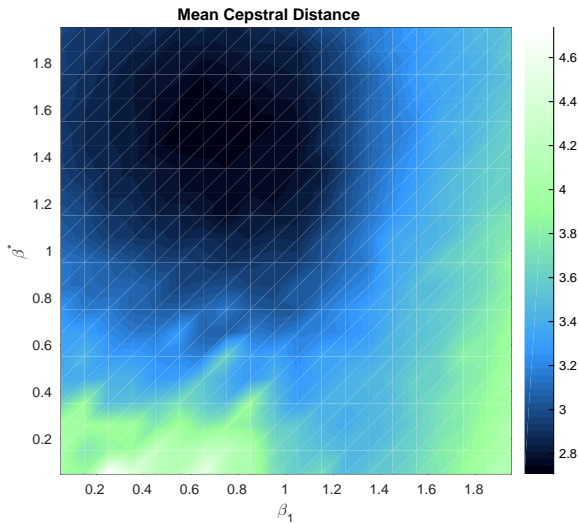}
\caption{Mean cepstral distance values obtained from a representation of a clean signal using a $\beta^*$ divergence, with a dictionary built from a reverberant version using $\beta_1$ . Smaller values correspond to better results.}
\label{fi:cdmap}
\end{figure}

Fig. \ref{fi:cdmap} depicts the resulting mean cepstral distance (over five trials over each of the five signals) as a function of the parameters $\beta_1$ and $\beta^*$. The minimizer is reached at (0.75,1.45), showing that $\beta_1 = 0.75$  is the best parameter choice for Stage 1 of Algorithm \ref{al:beta_der}. Note that this does not necessarily mean that $\beta_2 = 1.45$ is the best choice for the second stage of Algorithm \ref{al:beta_der}, since here we are minimizing $D_{\beta}(S||\hat{S})$ whereas the second step of the dereverberation method requires minimizing Equation (\ref{eq:functional}). 

It should be pointed out that functional (\ref{eq:functional}) is a generalization of a Bayesian approach (similar to the one in \cite{ibarrola2018bayesian}) if $U$ and $\nabla_t H$ are treated as random variables with exponential and normal \emph{a-priori} distributions, respectively. In fact, by choosing $\beta = 2$, the minimizer of (\ref{eq:functional}) corresponds to a \emph{maximum-a-posteriori} (MAP) estimator, given proper choices of the penalization parameters. Therefore, we have chosen  $\beta  = 2$ for Stage 2 of Algorithm \ref{al:beta_der}, which in fact was observed to lead to better results than $\beta  = 1.45$.

A few relevant conclusions can be derived by observing Fig \ref{fi:cdmap}. First, that the values of $(\beta_1,\beta^*)$ leading to the smallest cepstral distances are away from the diagonal, thus corroborating our original conjecture that using different parameter values for the learning and representation steps could lead to improved results. Furthermore, note that better results are obtained for values of $(\beta_1,\beta*)$ in the top left area. This most probably reflects the fact that small values of $\beta_1$ lead to dictionaries which take all the frequency range into account, whereas high values of $\beta^*$ promote fidelity on the high-energy zones of the represented spectrogram.

\subsection{Illustration}

\begin{figure*} [!t]
	\centering
\includegraphics[width=\textwidth]{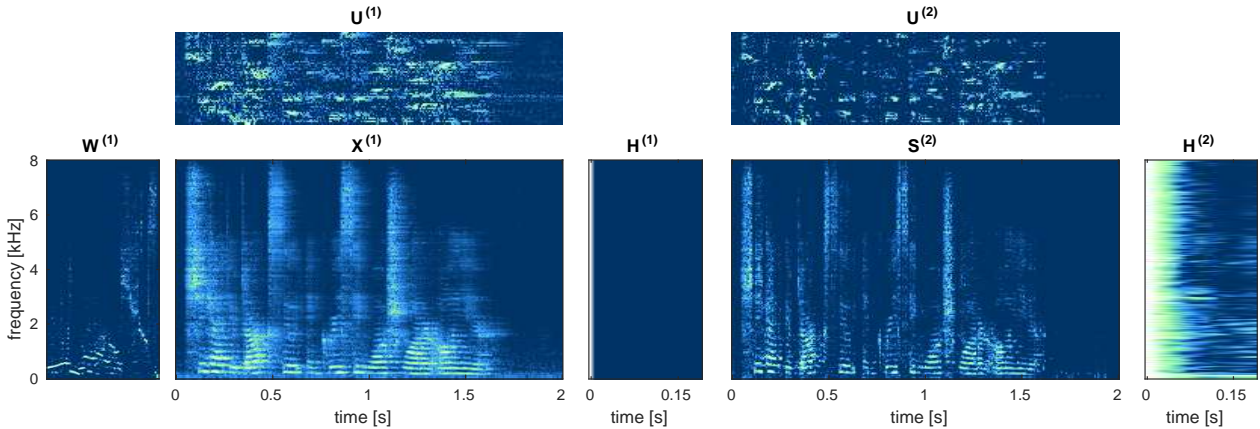}
\caption{Representation elements obtained with the proposed method. $W^{(1)}$, $U^{(1)}$, $H^{(1)}$, and  $S^{(1)}$=$W^{(1)}U^{(1)}$ are the matrices at the end of Stage 1, and  $U^{(2)}$, $H^{(2)}$, and  $S^{(2)}=W^{(1)}U^{(2)}$  are the matrices at the end of the dereverberation process. All the elements are in $\log$ scale, in amplitude.}
\label{fi:illustration}
\end{figure*}

Before beginning with the actual experiments we show how the method works by plotting the result obtained for just one signal. The signal corresponds to a female speaker pronouncing the sentence ``She had your dark suit in greasy wash water all year'', from the TIMIT database, recorded in an office room (Room 1, in Table \ref{tab:realrooms}) in real-life conditions, as specified in Section \ref{sec:recording_experiments}. All representation elements are depicted in Fig. \ref{fi:illustration}. It can be seen that at the end of Stage 1, a dictionary $W^{(1)}$ is built while reverberation is captured in the coefficient matrix $U^{(1)}$. In the second stage, reverberation is mostly represented by $H^{(2)}$, thus allowing the coefficients in $U^{(2)}$ to provide a good representation $S^{(2)}$ of the clean spectrogram $S$.

\subsection{Validation}

We have chosen two different settings for the validation experiments. The first one using simulations in order have a large number of trials available, and the second one using real recordings to guarantee the method is applicable in real-life conditions.

The model parameters used for all the experiments are detailed in Table \ref{tab:params}.

\begin{table}[!t]
\renewcommand{\arraystretch}{1.3}
\caption{Model parameters}
\label{tab:params}
\centering
\begin{tabular}{cccccc}
win. size & win. overl. & $J$ & $M$ & $\beta_1$  &$\beta_2$  \\
\hline
512 & 256 & 64 & 20 & 0.75 & 2 \\
\hline
\end{tabular}
\bigskip

\begin{tabular}{ccc}
 $\lambda^{(u)}_n$  & $\lambda_{k}^{(h)}$ & $\delta$  \\
\hline
$mean(Y)\times 10^{-3} $ &$0.3\|Y_k\|^2$  & $\|Y\|\times 10^{-3} $\\
\hline
\end{tabular}
\end{table}

In order to evaluate the performance of our method, comparisons against two state-of-the-art methods applicable under the same conditions were made. The first one was proposed in \cite{ibarrola2018bayesian}, and it has shown to perform quite well. The other one was proposed by Wisdom \emph{et al} in \cite{wisdom2014}, and showed an excellent performance in the Reverb Challenge (\cite{kinoshita2016summary}).

\subsubsection{Simulated experiments}

For the simulations, $110$ speech signals from the TIMIT database were taken, and made reverberant by convolution with artificial impulse responses. The artificial RIRs were generated varying the microphone positions and room dimensions, as specified in Table \ref{tab:rooms}. The reverberation time was set at either $450$[ms], $600$[ms] or $750$[ms], resulting in 27 different reverberation conditions, and hence a total of 2970 reverberant signals for testing.

\begin{table}[!t]
\caption{Simulated room settings}
\centering
\begin{tabular}{lccc}
\hline
 & Length & Width & Height \\
\hline
Room 1 dimensions  & 5.00 [m] & 4.00 [m] & 6.00 [m] \\
Room 2 dimensions  & 4.00 [m] & 4.00 [m] & 3.00 [m] \\
Room 3 dimensions  & 10.0 [m] & 4.00 [m] & 5.00 [m] \\
Source  position & 2.00 [m] & 3.50 [m] & 2.00 [m] \\
Microphone 1 position & 2.00 [m] & 1.50 [m] & 1.00 [m] \\
Microphone 2 position & 2.00 [m] & 2.00 [m] & 1.00 [m] \\
Microphone 3 position & 2.00 [m] & 2.00 [m] & 2.00 [m] \\
\hline
\end{tabular}
\label{tab:rooms}
\end{table}

\begin{figure}[!t]
\centering
\includegraphics[width=0.5\textwidth]{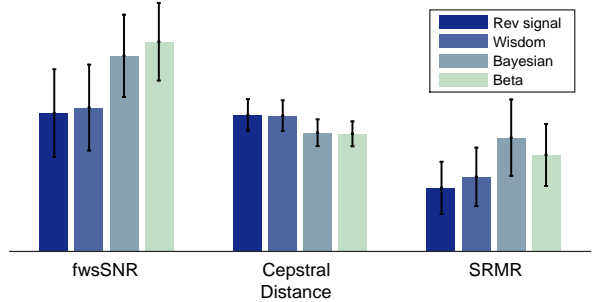}
\caption{Mean and standard deviation of performance measures for each method, using simulations. }
\label{fig:results_SIMS}
\end{figure}

Table \ref{tab:results_SIMS} and Fig. \ref{fig:results_SIMS} show the results obtained with each performance measure and each one of the methods. Note that our proposed method (labeled ``Beta'') outperforms ($p<0.01$) the other two in terms of fwsSNR and cepstral distance, but not the Bayesian (\cite{ibarrola2018bayesian}) in terms of SRMR. However, taking into account that SRMR quantifies the extent to which  a signal ``seems'' reverberant, but not how much such a restoration resembles the corresponding clean signal, it should only be considered as a complement to the other two measures.

\begin{table}[!t]
\caption{Mean and standard deviation (between parenthesis) of performance measures for each method, using simulations. Best results are shown in boldface.}
\centering
\begin{tabular}{ lccc }
\hline
Measure & fwsSNR & Cepstral Dist. & SRMR  \\  
\hline
Reverberant & 5.377 (1.70) & 5.308 (0.61) & 2.470 (1.01) \\ 
Wisdom & 5.593 (1.67) & 5.279 (0.60) & 2.898 (1.14) \\ 
Bayesian & 7.604 (1.60) & 4.614 (0.52) & \textbf{4.423} (1.48) \\ 
Beta & \textbf{8.153} (1.51) & \textbf{4.573} (0.48) & 3.751 (1.21) \\ 
\hline
\end{tabular}\normalsize
\label{tab:results_SIMS}
\end{table}

\subsubsection{Experiments using recordings} \label{sec:recording_experiments}

In order to test whether our method works in real-life situations, we made recordings in two of our own office rooms, during standard office hours and with air conditioners and computers left on. The offices' dimensions are shown in Table \ref{tab:realrooms}, along with the speaker and microphone positions. The reverberation times of the rooms turned out to be of $460$[ms] in Room 1 and of $440$[ms] in Room 2, as measured using sine sweeps (\cite{farina2007advancements}). Four speakers (two male and two female) were randomly selected from the TIMIT database, and 10 speech signals from each were recorded in each room, with a sampling frequency of $16$[kHz].

As it is customary, the clean speech sources had their low-frequency components filtered out. Hence, we pre-processed our reverberant recordings using a $5000$ tap FIR high-pass filter with cut-off frequency of $30$[Hz] to mitigate the low frequency noise. For the comparisons to be fair, all the methods were tested after this pre-processing was made.

In order to better cope with the noise, the penalization parameters for $U$ were reset to $\lambda^{(u)}_n = \frac{mean(Y)}{\|U^1_n\|_1}\times 10^{-1}$, where $U^1_n$ is the $n$-th column of $U$ as estimated at the end of Stage 1 of Algorithm \ref{al:beta_der}. This prevents the model from attempting to represent ambient noise during speech silences.

\begin{table}[!t]
\caption{Office rooms settings}
\centering
\begin{tabular}{lccc}
\hline
 & Length & Width & Height \\
\hline
Room 1 dimensions  			& 4.15 [m] & 3.00 [m] & 3.00 [m] \\
Source  1 position 				& 3.60 [m] & 1.50 [m] & 1.50 [m] \\
Microphone 1 position 	& 1.10 [m] & 1.50 [m] & 1.50 [m] \\
\hline
Room 2 dimensions  		   & 5.85 [m] & 4.55 [m]  & 3.00 [m] \\
Source  2 position 				& 1.10 [m] & 1.50 [m] & 1.50 [m] \\
Microphone 2 position 	& 1.10 [m] & 4.00 [m] & 1.50 [m] \\
\hline
\end{tabular}
\label{tab:realrooms}
\end{table}

\begin{figure}[!t]
\centering
\includegraphics[width=0.5\textwidth]{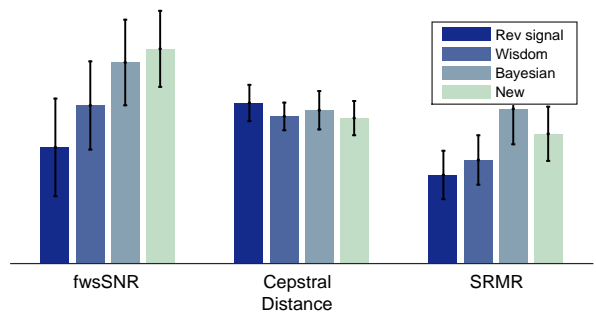}
\caption{Mean and standard deviation of performance measures for each method, using recordings.}
\label{fig:results_RECS}
\end{figure}

Results are depicted in Table \ref{tab:results_rec} and illustrated in Figure \ref{fig:results_RECS}. Once again, we see that our proposed method outperforms the  others in terms of the fwsSNR, but loses to the Bayesian in terms of SRMR. As for the cepstral distance, the improvement between our proposed method and Wisdom's is the only one not reaching statistical significance ($p>0.01$).

\begin{table}[!t]
\caption{Mean and standard deviation (between parenthesis) of performance measures for each method. Best results are shown in boldface.}
\centering
\begin{tabular}{ l c c c }
\hline
Measure & fwsSNR & Cepstral Dist. & SRMR  \\  
\hline
Reverberant & 3.613 (1.52) & 4.994 (0.56) & 2.756 (0.75) \\ 
Wisdom & 4.917 (1.37) & 4.577 (0.43) & 3.222 (0.77) \\ 
Bayesian & 6.254 (1.33) & 4.769 (0.60) & \textbf{4.809 }(1.10) \\ 
Beta & \textbf{6.678} (1.18) & \textbf{4.524} (0.53) & 4.036 (0.84) \\
\hline
\end{tabular}\normalsize
\label{tab:results_rec}
\end{table}

\section{Conclusions}

In this work, a new blind, single channel dereverberation method in the time-frequency domain that makes use of variable $\beta$-divergence as a cost function was presented and tested. The method comprises two stages: one for learning the spectral structure into a dictionary, and a second one for using such a dictionary to build an accurate representation by means of a convolutive NMF model. The corresponding algorithm for implementing the method was introduced and tested. Additionally, a method for finding an optimal learning divergence was introduced.

Results show that the proposed method improves restoration quality with respect to state-of-the-art methods, as measured by the fwsSNR and cepstral distance. Improvement in regard to SRMR is only partial, but being this a non-intrusive measure, that is not too much of a drawback.

There is certainly much room for improvement. For instance, exploring the use of penalization terms at the learning stage and other ways of enhancing the quality of the dictionary, as well as generating atoms for specifically modeling (and then removing) noise and incorporating specific initialization methods. All this is subject of future study.

Finally, although our method is constructed for a blind setting, it is worth noting that it can be easily adapted to be supervised by modifying the learning stage, provided speaker information is available.

\section*{Acknowledgments}

This research was funded by ANPCyT under projects PICT 2014-2627 and PICT 2015-0977, by UNL under projects CAI+D 50420150100036LI, CAI+D 50020150100059LI,  CAI+D 50020150100055LI and CAI+D 50020150100082LI.



\bibliographystyle{IEEEtran}
 \bibliography{ref_der}

\begin{thebibliography}{10}
\providecommand{\url}[1]{#1}
\csname url@samestyle\endcsname
\providecommand{\newblock}{\relax}
\providecommand{\bibinfo}[2]{#2}
\providecommand{\BIBentrySTDinterwordspacing}{\spaceskip=0pt\relax}
\providecommand{\BIBentryALTinterwordstretchfactor}{4}
\providecommand{\BIBentryALTinterwordspacing}{\spaceskip=\fontdimen2\font plus
\BIBentryALTinterwordstretchfactor\fontdimen3\font minus
  \fontdimen4\font\relax}
\providecommand{\BIBforeignlanguage}[2]{{%
\expandafter\ifx\csname l@#1\endcsname\relax
\typeout{** WARNING: IEEEtran.bst: No hyphenation pattern has been}%
\typeout{** loaded for the language `#1'. Using the pattern for}%
\typeout{** the default language instead.}%
\else
\language=\csname l@#1\endcsname
\fi
#2}}
\providecommand{\BIBdecl}{\relax}
\BIBdecl

\bibitem{yun_multilingual_2014}
S.~Yun, Y.~J. Lee, and S.~H. Kim, ``Multilingual speech-to-speech translation
  system for mobile consumer devices,'' \emph{IEEE Transactions on Consumer
  Electronics}, vol.~60, no.~3, pp. 508--516, 2014.

\bibitem{vignolo_feature_2016}
L.~D. Vignolo, S.~R.~M. Prasanna, S.~Dandapat, H.~L. Rufiner, and D.~H. Milone,
  ``Feature optimisation for stress recognition in speech,'' \emph{Pattern
  Recognition Letters}, vol.~84, pp. 1--7, 2016.

\bibitem{sarikaya2016overview}
R.~Sarikaya, P.~A. Crook, A.~Marin, M.~Jeong, J.-P. Robichaud, A.~Celikyilmaz,
  Y.-B. Kim, A.~Rochette, O.~Z. Khan, X.~Liu \emph{et~al.}, ``An overview of
  end-to-end language understanding and dialog management for personal digital
  assistants,'' in \emph{Spoken Language Technology Workshop (SLT), 2016
  IEEE}.\hskip 1em plus 0.5em minus 0.4em\relax IEEE, 2016, pp. 391--397.

\bibitem{tashev2009sound}
I.~J. Tashev, \emph{Sound capture and processing: practical approaches}.\hskip
  1em plus 0.5em minus 0.4em\relax John Wiley \& Sons, 2009.

\bibitem{huang2001spoken}
X.~Huang, A.~Acero, H.-W. Hon, and R.~Reddy, \emph{Spoken language processing:
  A guide to theory, algorithm, and system development}.\hskip 1em plus 0.5em
  minus 0.4em\relax Prentice hall PTR Upper Saddle River, 2001, vol.~95.

\bibitem{wisdom2014}
S.~Wisdom, T.~Powers, L.~Atlas, and J.~Pitton, ``Enhancement of reverberant and
  noisy speech by extending its coherence,'' in \emph{Proceedings of REVERB
  Challenge Workshop}, 2014, pp. 1--8.

\bibitem{ibarrola2018bayesian}
F.~Ibarrola, L.~Di~Persia, and R.~Spies, ``A bayesian approach to convolutive
  nonnegative matrix factorization for blind speech dereverberation,''
  \emph{Signal Processing}, vol. 151, pp. 89--98, 2018.

\bibitem{smaragdis2004}
P.~Smaragdis, ``Non-negative matrix factor deconvolution; extraction of
  multiple sound sources from monophonic inputs,'' \emph{Proceedings of the 5th
  Conference on Independent Component Analysis and Blind Signal Separation},
  pp. 494--499, 2004.

\bibitem{mohammadiha2015joint}
N.~Mohammadiha, P.~Smaragdis, and S.~Doclo, ``Joint acoustic and spectral
  modeling for speech dereverberation using non-negative representations,'' in
  \emph{Acoustics, Speech and Signal Processing (ICASSP), 2015 IEEE
  International Conference on}.\hskip 1em plus 0.5em minus 0.4em\relax IEEE,
  2015, pp. 4410--4414.

\bibitem{avargel2007system}
Y.~Avargel and I.~Cohen, ``System identification in the short-time {Fourier}
  transform domain with crossband filtering,'' \emph{IEEE Transactions on
  Audio, Speech, and Language Processing}, vol.~15, no.~4, pp. 1305--1319,
  2007.

\bibitem{yegnanarayana1998}
B.~Yegnanarayana, P.~S. Murthy, C.~Avenda{\~n}o, and H.~Hermansky,
  ``Enhancement of reverberant speech using lp residual,'' in \emph{Acoustics,
  Speech and Signal Processing, 1998. Proceedings of the 1998 IEEE
  International Conference on}, vol.~1.\hskip 1em plus 0.5em minus 0.4em\relax
  IEEE, 1998, pp. 405--408.

\bibitem{kameoka2009}
H.~Kameoka, T.~Nakatani, and T.~Yoshioka, ``Robust speech dereverberation based
  on non-negativity and sparse nature of speech spectrograms,'' in \emph{2009
  IEEE International Conference on Acoustics, Speech and Signal Processing},
  2009, pp. 45--48.

\bibitem{fevotte2009nonnegative}
C.~F{\'e}votte, N.~Bertin, and J.-L. Durrieu, ``Nonnegative matrix
  factorization with the itakura-saito divergence: With application to music
  analysis,'' \emph{Neural computation}, vol.~21, no.~3, pp. 793--830, 2009.

\bibitem{kompass2007generalized}
R.~Kompass, ``A generalized divergence measure for nonnegative matrix
  factorization,'' \emph{Neural computation}, vol.~19, no.~3, pp. 780--791,
  2007.

\bibitem{de2015modeling}
E.~De~Sena, N.~Antonello, M.~Moonen, and T.~Van~Waterschoot, ``On the modeling
  of rectangular geometries in room acoustic simulations,'' \emph{IEEE/ACM
  Transactions on Audio, Speech and Language Processing (TASLP)}, vol.~23,
  no.~4, pp. 774--786, 2015.

\bibitem{ratnam2003blind}
R.~Ratnam, D.~L. Jones, B.~C. Wheeler, W.~D. O’Brien~Jr, C.~R. Lansing, and
  A.~S. Feng, ``Blind estimation of reverberation time,'' \emph{The Journal of
  the Acoustical Society of America}, vol. 114, no.~5, pp. 2877--2892, 2003.

\bibitem{lee2001}
D.~D. Lee and H.~S. Seung, ``Algorithms for non-negative matrix
  factorization,'' in \emph{Advances in Neural Information Processing Systems},
  2001, pp. 556--562.

\bibitem{fevotte2011algorithms}
C.~F{\'e}votte and J.~Idier, ``Algorithms for nonnegative matrix factorization
  with the $\beta$-divergence,'' \emph{Neural computation}, vol.~23, no.~9, pp.
  2421--2456, 2011.

\bibitem{choi2005blind}
S.~Choi, A.~Cichocki, H.-M. Park, and S.-Y. Lee, ``Blind source separation and
  independent component analysis: A review,'' \emph{Neural Information
  Processing-Letters and Reviews}, vol.~6, no.~1, pp. 1--57, 2005.

\bibitem{zue1990timit}
V.~Zue, S.~Seneff, and J.~Glass, ``Speech database development at {MIT}:
  {TIMIT} and beyond,'' \emph{Speech Communication}, vol.~9, no.~4, pp.
  351--356, 1990.

\bibitem{hu2008}
Y.~Hu and P.~C. Loizou, ``Evaluation of objective quality measures for speech
  enhancement,'' \emph{IEEE Transactions on Audio, Speech, and Language
  Processing}, vol.~16, no.~1, pp. 229--238, 2008.

\bibitem{falk2010}
T.~H. Falk, C.~Zheng, and W.-Y. Chan, ``A non-intrusive quality and
  intelligibility measure of reverberant and dereverberated speech,''
  \emph{IEEE Transactions on Audio, Speech, and Language Processing}, vol.~18,
  no.~7, pp. 1766--1774, 2010.

\bibitem{kinoshita2016summary}
K.~Kinoshita, M.~Delcroix, S.~Gannot, E.~A. Habets, R.~Haeb-Umbach,
  W.~Kellermann, V.~Leutnant, R.~Maas, T.~Nakatani, B.~Raj \emph{et~al.}, ``A
  summary of the reverb challenge: state-of-the-art and remaining challenges in
  reverberant speech processing research,'' \emph{EURASIP Journal on Advances
  in Signal Processing}, vol. 2016, no.~1, p.~7, 2016.

\bibitem{farina2007advancements}
A.~Farina, ``Advancements in impulse response measurements by sine sweeps,'' in
  \emph{Audio Engineering Society Convention 122}.\hskip 1em plus 0.5em minus
  0.4em\relax Audio Engineering Society, 2007.

\end{thebibliography}

\end{document}